\documentclass[10pt,conference]{IEEEtran}
\IEEEoverridecommandlockouts
\usepackage{cite}
\usepackage{amsmath,amssymb,amsfonts}
\usepackage{algorithmic}
\usepackage{graphicx}
\usepackage{textcomp}
\usepackage{xcolor}
\def\BibTeX{{\rm B\kern-.05em{\sc i\kern-.025em b}\kern-.08em
    T\kern-.1667em\lower.7ex\hbox{E}\kern-.125emX}}

\usepackage[linesnumbered,ruled,vlined]{algorithm2e}
\usepackage{hyperref}
\usepackage{tabulary}
\usepackage{tablefootnote}

\usepackage{xurl} %

\usepackage{float}

\usepackage{graphics}
\graphicspath{ {./figs/} }

\usepackage{amsthm}

\usepackage{booktabs}

\usepackage{paralist}

\usepackage{makecell}
\usepackage{multirow}

\usepackage{enumitem}

\usepackage{pgfplots}
\pgfplotsset{compat=1.18}
\usepackage{pgfplotstable}

\usepackage[most]{tcolorbox}
\tcbset{
  colback=green!5!white,
  colframe=green!85!black,
  boxsep=4pt,                 %
  left=0pt, right=0pt,        %
  top=0pt, bottom=0pt,        %
  enhanced,
  sharp corners               %
}

\usepackage{soul}

\newcommand{\step}[1]{%
  \tikz[baseline=(char.base)]{
    \node[shape=circle,fill=black,text=white,inner sep=0pt,minimum size=12pt,align=center] (char) {\sffamily\small #1};
  }%
}

\usetikzlibrary{calc, shapes, arrows.meta, positioning}

\definecolor{color-up-to-date}{HTML}{8CC5E3}
\definecolor{color-tood}{HTML}{2066A8}
\definecolor{color-pfet-tood}{HTML}{F72B8F}

\SetCommentSty{mycommfont}

\begin{document}

\title{
Relative Positioning Based Code Chunking Method For Rich Context Retrieval In Repository Level Code Completion Task With Code Language Model
}

\author{
\IEEEauthorblockN{Imranur Rahman\IEEEauthorrefmark{3} and Md Rayhanur Rahman\IEEEauthorrefmark{2}}
\IEEEauthorblockA{\IEEEauthorrefmark{3}North Carolina State University, \IEEEauthorrefmark{2}The University of Alabama\\
\IEEEauthorrefmark{3}irahman3@ncsu.edu, \IEEEauthorrefmark{2}mdrayhanur.rahman@ua.edu}
}

\maketitle

\begin{abstract}
Code completion can help developers improve efficiency and ease the development lifecycle.
Although code completion is available in modern integrated development environments (IDEs), research lacks in determining what makes a good context for code completion based on the information available to the IDEs for the large language models (LLMs) to perform better.
In this paper, we describe an effective context collection strategy to assist the LLMs in performing better at code completion tasks.
The key idea of our strategy is to preprocess the repository into smaller code chunks and later use syntactic and semantic similarity-based code chunk retrieval with relative positioning.
We found that code chunking and relative positioning of the chunks in the final context improve the performance of code completion tasks.
\end{abstract}


\section{Introduction}
Code completion is one of the most aspired features for developers in an integrated development environment (IDE)~\cite{semenkin_context_nodate}.
Prominent IDEs already have such a feature to assist developers in increasing their productivity.
For example, two of the most popular IDEs, Visual Studio Code~\cite{copilot_in_vs_code} and IntelliJ~\cite{intellij_code_completion}, have provided code completion features to developers.
However, not all the code completions generated by the state-of-the-art code completion systems are helpful for the users~\cite{sun_dont_2024}.
The current code completion feature lacks in several dimensions: (1) complexity and quality of the code completion, (2) use of local vs cloud large language models (LLMs), and (3) identification of relevant context to help the LLMs in the code completion task~\cite{semenkin_full_2025}.

For the AI-enabled systems, the code completion quality heavily depends on how well the model understands the surrounding code, the context, and how well the model understands users' intent.
Sapronov et al.~\cite{sapronov_pretraining_2025} found that the context provided to the model plays an important role in the quality of the code generated by LLM.
To improve the broad understanding of the context collection, JetBrains Research organized a workshop co-located with the Automated Software Engineering (ASE 2025) conference and asked the community to come up with an effective context collection strategy.
The objective of the competition is to create a context collection strategy that supplements the provided completion points with useful information from across the whole repository.

At the problem setup, we were given: (1) the full source code of a repository, (2) recently used files by the user, (3) the completion file in which the code is being written, and (4) a prefix and a suffix of a code snippet.
The LLM is supposed to `fill in the middle' between the prefix and the suffix, but the participants did not have access to the LLM.
The task for the participants of the competition was to provide context based on the information provided so that the LLM performs better in the code completion task with the provided context.

We first preprocess the whole repository into code chunks and later retrieve relevant code chunks for the input: prefix and suffix.
\textbf{
Our key idea was to attach \textit{next} code chunks with the retrieved similar code chunks for the \textit{prefix} as context, since the \textit{next} code chunks represent what the user has already written \textit{after} the retrieved similar code.
Similarly, we attach \textit{prev} code chunks with the retrieved similar code chunks for the \textit{suffix} as context, since the \textit{prev} code chunks represent what the user has already written \textit{before} the retrieved similar code.
}
We are calling this \textit{relative positioning}.
Our final context combines (1) the completion file, (2) the recently opened files by the user, (3) top-k similar code chunks to prefix alongside their \textit{next} code chunks, and (4) top-k similar code chunks to suffix alongside their \textit{prev} code chunks.
With this solution, we (team name: WSPR\_NCSU) have received 0.660 as the average chrF score in Kotlin and 0.636 as the average in the Python track.
Our solution won third place (Bronze) in the Kotlin track and scored fourth place in the Python track of the competition.
We open source our implementation in \href{https://github.com/imranur-rahman/ase2025-starter-kit/blob/main/baselines.py}{https://github.com/imranur-rahman/ase2025-starter-kit/blob/main/baselines.py}.

\section{Solution Overview}

\subsection{Preprocessing}
At the preprocessing step, we chunk each source file from the repository and store relevant information in databases for later retrieval.
First, we split each source file into \textit{n} line chunks where there is a \textit{m} line overlap in two consecutive chunks (Step \step{1}).
We also keep track of the \textit{prev} and \textit{next} pointers of each chunk within a file.
If a source file consists of fewer than \textit{m} lines, we consider the file as one chunk with no \textit{prev} or \textit{next} pointers.
After chunking, we store each chunk in a chunk database with the \textit{prev} and \textit{next} pointers for later syntactic retrieval in the later stage (Step \step{2}).
At the same time, we compute the vector embeddings for each of the chunks using an embedding model (Step \step{3}).
We used \texttt{sentence-transformers/all-MiniLM-L6-v2} as the embedding model from HuggingFace since the model is small enough for faster embedding generation (22.7M parameters).
This sentence-transformers model maps sentences and paragraphs to a 384-dimensional dense vector space and can be used for tasks like clustering or semantic search~\cite{embedding_model}.
This embedding model is general-purpose for NLP tasks, but not specifically designed for code-related tasks.
The tokenizer used by \texttt{all-MiniLM-L6-v2} is for natural language.
Using an embedding model that is pretrained specifically for code-related tasks should improve the tokenization and, in turn, the embedding generation, which, in turn, should improve the overall performance of our system.
However, due to our limited computing resources, we opted for \texttt{all-MiniLM-L6-v2}.
After generating the embedding, we use a vector database for storing the embedding.

\subsection{Retriever}
We use an Ensemble Retriever~\cite{ensemble_retriever} from LangChain.
Ensemble Retriever supports combining results from multiple retrievers.
The ensemble retriever reranks the results of the constituent retrievers based on the Reciprocal Rank Fusion algorithm~\cite{cormack2009reciprocal}.
In our solution, we employ a hybrid approach of using both \textit{syntactic similarity} and \textit{semantic similarity} when looking for similar code chunks.
For syntactic similarity, we use a BM25 retriever, which retrieves chunks from the chunk database.
BM25~\cite{bm25} is a sparse retriever that is good at finding relevant documents based on keywords.
For the semantic similarity, we use FAISS (Facebook AI Similarity Search)~\cite{douze2024faiss,johnson2019billion}, which is designed for efficient similarity search and clustering of dense vectors.
FAISS is a dense retriever that is good at finding relevant documents based on semantic similarity.

The ensemble retriever has configurable parameters (e.g., how many results to be retrieved, corresponding weights for each of the individual retrievers).
We configured the ensemble retriever to return top-30 code chunks after reranking.
The relative weights for BM25 and FAISS retrievers were 0.2 and 0.8.
With this relative weight, we put 4 times more emphasis on the semantic similarity than the syntactic similarity since our intuition was that semantic similarity is more important for code-related tasks than syntactic similarity.

\subsection{Context Collection}
\label{sec:context-collection}
After preprocessing the repository and configuring the retriever, we begin our context collection step.
For each of the inference steps, we have a prefix and a suffix of a code snippet, and the task is to provide relevant information for the LLM to fill-in-the-middle.
First, we feed the prefix to the ensemble retriever to retrieve the top-k similar (both syntactic and semantic) code chunks from the database (Step \step{4}).
We then use the chunk database to retrieve the \textit{next} chunks for each of the top-k similar code chunks to prefix (Step \step{5}).
Our intuition was that for the LLM to fill-in-the-middle, the \textit{next} code chunks for the top-k similar code chunks to the prefix might be more useful.
Similarly, we feed the suffix to the ensemble retriever to retrieve the top-k similar (both syntactic and semantic) code chunks from the database (Step \step{6}).
Then, we use the chunk database to retrieve the \textit{prev} chunks for each of the top-k similar code chunks to suffix (Step \step{7}).
Again, the intuition we had was that for the LLM to fill-in-the-middle, the \textit{prev} code chunks for the top-k similar code chunks to the suffix might be more useful.
\textbf{Our key idea} was using \textit{next} code chunks of the top-k similar code chunks for prefix and \textit{prev} code chunks of the top-k similar code chunks for suffix.

\subsection{Final Context}
\begin{figure*}[t]
    \centering
    \includegraphics[width=\linewidth]{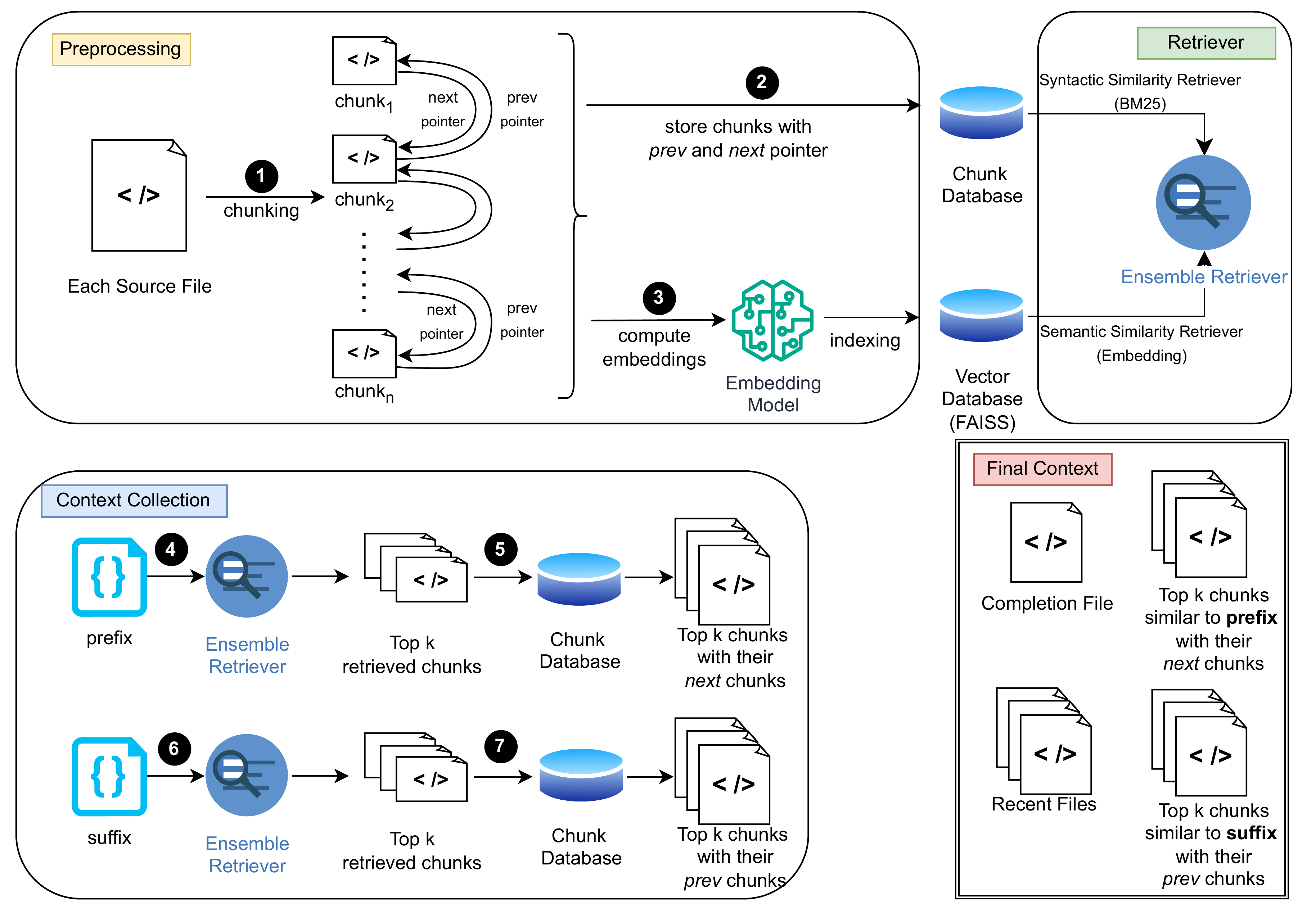}
    \caption{High-level overview of our solution.}
    \label{fig:overview}
\end{figure*}

Our final context generation for each of the inference steps is illustrated in Figure~\ref{fig:overview}.
The first part of the final context is the completion file.
Our intuition was that the completion file where the code snippet is being written might be the most useful for the LLM since the user is most likely to write the code complementary to the completion file.
The complementary code could be adding a feature, extending one function for a different purpose, or reusing the already present functions.
In addition to that, the LLM needs to know what variables and functions are available in the completion file to effectively reuse them even when it understands what is about to be written.
The second part of the final context is the recent files that the user has opened.
Since the recent files are provided alongside the prefix and suffix as an input, and the baseline provided from the organizers was \texttt{recent}, we opted for keeping the recent files as a part of the final context.
The intuition behind having the recent files as the final context was that if the user is adding a feature to the completion file, adding a functionality for another file, or reusing the functionalities from another file, the recent files are the files most likely ones that could provide the context to the LLM what is present in the other file or what could be added.
The next two parts are the outputs of our context collection step from Section~\ref{sec:context-collection}: top-k similar chunks to prefix with their \textit{next} chunks, top-k similar chunks to suffix with their \textit{prev} chunks.

\section{Incremental Progression To The Final Solution}
We have tried multiple different approaches before coming to our final solution.
Initially, we tried statically extracting the variable names and function names present in a file and attaching them alongside the file.
The intuition was that LLMs could have difficulties in parsing the file to know what variables and functions are already present in the file that the LLM can reuse.
Also, we used a JSON-formatted object to pass these information to the LLM inside the code context since the prompt formatting has an impact on the performance of the LLM~\cite{he2024doespromptformattingimpact}.
Izadi et al.~\cite{izadi_language_2024} found that incorrect variable, incorrect function, incorrect literal, and incorrect type were the top-4 reasons why code completion fails in real-world settings.
However, in this particular task, a structured context with a correct variable name and a correct function name did not provide any additional performance benefit in our evaluation.

Next, we tried one type of prompt injection to ask the LLM whether it needs additional context to fill-in-the-middle part of the prefix and suffix.
Since LLMs do not have a good understanding of what is instruction and what is data, we can add a question to the context, ``\textit{Do you need additional context to fill up the code? If yes, use the provided context below; otherwise, discard the additional context.}''
This idea came from RepoFormer~\cite{wu_repoformer_2024}, where the authors asked a critical question, ``Should we always perform retrieval augmentation?''
To answer this question, they have an additional \textit{Retrieval Decision} step where they prompt the LLM to show its confidence score.
If the LLM is confident, they do not provide the additional context; otherwise, they do.
The intuition behind this approach was that there might be some cases where the code LM already knows the answer without retrieval~\cite{kadavath_language_2022}, and the code completion question does not depend on cross-file information, and thus retrieval is likely uninformative.
Since we do not have access to the model before for this \textit{Retrieval Decision} step, we opted for prompt injecting the question into the context.
However, this approach did not provide any additional benefit in accuracy.
We also tried having a locally hosted LLM (\texttt{Qwen2.5-coder}) to avoid prompt injecting and directly ask the question to the LLM.
Our intuition was that if our locally hosted LLM is confident in the fill-in-the-middle process, the LLMs used in the competition would be confident as well.
We only added the context when our locally hosted LLM was less confident than a confidence threshold.
However, we did not see any performance improvement with this \textit{selective retrieval} strategy.
There were two issues with this approach: (1) not all LLMs are similar; so having a good confidence score in our locally hosted LLM does not necessarily mean the used LLMs in the competition (Mellum by JetBrains, Codestral by Mistral AI, Qwen2.5-Coder by Alibaba Cloud), and (2) since we had to do inference locally for each of the code completion task, it was becoming computationally expensive and time consuming which was over the time limit set by the competition.

After trying the above two approaches, we were wondering if there was a simple approach that does not involve any kind of prompt injection or inference on a locally hosted LLM.
We started thinking about what information would be really helpful for the LLM for the code completion tasks, and that can be computed with fewer computing resources and within a short time.
\textbf{That is when we came up with the idea of chunking the code and attaching not only the similar code chunks but also the \textit{next} and \textit{prev} code chunks based on the similar code chunks of prefix and suffix, respectively.}
The organizers have provided another baseline where they have provided the BM25 retrieval solution, and our approach builds on top of the syntactic similarity retrieval approach of BM25.
This is the solution we described in the paper, and we have achieved third place in the `Kotlin' track of the competition.

Throughout the time of the competition, our goal was to create a unified solution that is language agnostic, since having a language agnostic solution was one of the motivations for the organizers to organize this competition and workshop.
Since our code chunking strategy was naive and static, it may break the relatively coherent code into multiple code chunks.
So we tried a dynamic chunking strategy based on the abstract syntax tree (AST) of the code to have a hierarchical structure to ensure that coherent code stays in one chunk.
For this dynamic chunking, we did not have a strict number of lines for each chunk.
However, we could only partially implement this strategy with the \texttt{ast} module of Python, but not for Kotlin. 
We were not aware of the \texttt{TreeSitter} module, which is language agnostic and provides a tree-like representation of the source code of a program, capturing its structural elements and their relationships.
We envision that such a dynamic chunking strategy would provide us with a better performance in the code completion task with the help of a structured representation of the source code (e.g., abstract syntax tree, control flow graph, code property graph).

\section{Discussion And Conclusion}
Context collection is a challenging problem for AI-enabled code completion tasks.
In our solution, we opted for a one-phase preprocessing step on the repository.
However, the user may have multiple repositories in a real-world setting, and learning the code writing convention of the user might be helpful for the LM for code completion tasks.
Other repository-level information might be helpful for the code completion tasks, e.g., inter-file and cross-file dependency~\cite{ding_cocomic_2023}, call graph.
We expect that AI-enabled systems might benefit from this type of hierarchical representation of the repository.
As an extension of our work, an ablation study might provide deeper insight into how much improvement each of the components of our final context provides to the final outcome.
In addition, understanding how other factors, e.g., different chunk size, top-k values, influence the models performance might be very interesting.
With all these, we should also consider the memory footprint and time needed to aggregate the relevant context as metrics, to evaluate the context collection strategy, since if the code completion takes huge memory, the code completion might lose its attraction to users.

\section{Team Roles And Collaboration Process}
Imranur Rahman is the lead author and worked on idea generation, implementation, and testing. Md Rayhanur Rahman is the advisor who oversaw the whole process and helped with the preparation of this manuscript.

\bibliographystyle{IEEEtran}
\bibliography{websites}

\end{document}